\documentclass{article}
\usepackage[utf8]{inputenc}
\usepackage{booktabs}
\usepackage{verbatim}
\usepackage{amsmath}
\usepackage{amssymb}
\usepackage{amsfonts}
\usepackage{parskip}
\usepackage{mathtools}
\usepackage{caption}
\usepackage{subcaption}
\usepackage{graphicx}
\usepackage[normalem]{ulem}
\usepackage{pgfplots}
\usepackage{optidef}
\usepackage{float}
\usepackage[font=small,labelfont=bf]{caption}
\pgfplotsset{compat=newest}
\pgfplotscreateplotcyclelist{line styles}{
    black,solid\\
}
\newcommand*\GnuplotDefs{
    set samples 100;
    cdfn(z) = 0.5 * ( 1 + erf( z/sqrt(2) ));
    phi(xi) = exp(-0.5*xi^2)/sqrt(2*pi);
    tpdfn(x,mu,sd,a,b) = (1/sd) * (phi((x - mu)/sd))/( cdfn((b - mu)/sd) - cdfn((a - mu)/sd) );
}
\usepackage{tikz}
\usetikzlibrary{decorations.pathreplacing, calligraphy}

\usepackage[
doi=false,isbn=false,url=false,eprint=false,
backend=biber,
style=numeric,
sorting=none,
giveninits=true,
maxbibnames=3
]{biblatex}
\renewbibmacro{in:}{}
\AtEveryBibitem{%
  \clearlist{language}%
  \clearfield{pages}
  \clearfield{note}
  \clearfield{month}
}
\title{Implications of Clinical Target Distribution Weighted Radiotherapy Optimization}

\author{Ivar Bengtsson%
         \thanks{Optimization and Systems Theory, Department of
             Mathematics, KTH Royal Institute of Technology, SE-100 44
             Stockholm, Sweden (\texttt{ivarben@kth.se}, \texttt{andersf@kth.se}).}
        \thanks{RaySearch Laboratories AB, SE-104 30
             Stockholm, Sweden (\texttt{afred@raysearchlabs.com}).}
  \and Anders Forsgren\footnotemark[1]%
  \and Albin Fredriksson\footnotemark[2]%
  }
  
\date{}

\begin{document}

\maketitle

\begin{abstract}
    Delineating and planning with respect to regions suspected to contain microscopic tumor cells is an inherently uncertain task in radiotherapy. The recently proposed \textit{clinical target distribution} (CTD) is an alternative to the conventional \textit{clinical target volume} (CTV), with initial promise. Previously, using the CTD in planning has primarily been evaluated in comparison to a conventionally defined CTV. We propose to compare the CTD approach against CTV margins of various sizes, dependent on the threshold at which the tumor infiltration probability is considered relevant. First, a theoretical framework is presented, concerned with optimizing the trade-off between the probability of sufficient target coverage and the penalties associated with high dose. From this framework we derive conventional CTV-based planning and contrast it with the CTD approach. The approaches are contextualized further by comparison with established methods for managing geometric uncertainties. Second, for both a one- and a three-dimensional phantom, we compare a set of CTD plans created by varying the target objective function weight against a set of plans created by varying both the target weight and the CTV margin size. The results show that CTD-based planning gives slightly inefficient trade-offs between the evaluation criteria for a case in which near-minimum target dose is the highest priority. However, in a case when sparing a proximal organ at risk is critical, the CTD is better at maintaining sufficiently high dose toward the center of the target. We conclude that CTD-based planning is a computationally efficient method for planning with respect to delineation uncertainties, but that the inevitable effects on the dose distribution should not be disregarded.
\end{abstract}

\small \textbf{Keywords:} \textit{Radiotherapy optimization, clinical target distribution, target delineation uncertainty.}

\section{Introduction}

Accurate delineation of \textit{regions of interest} (ROIs) is a vital part of radiation therapy treatment planning. Conventionally, delineation is performed by an experienced radiation oncologist in a manual and time-consuming process. Yet, it is sometimes referred to as the weakest link in the radiotherapy chain, and many studies suggest that there is considerable inter-observer variability between clinicians, especially with regard to the \textit{clinical target volume} (CTV) \cite{unkelbach_role_2020, weiss_impact_2003, fiorino_grand_2020}. The CTV is defined in ICRU report 50 \cite{international_commission_on_radiation_units_and_measurements_prescribing_1993} as the volume suspected to contain microscopic tumour infiltration with clinically relevant probability. A CTV delineated by a clinician will thus depend largely on two factors: the perceived probability distribution of the tumorous volume and the threshold at which probability of tumor presence is considered relevant. The investigations in this work address the ambiguity associated with the second factor. We will thus assume a known probability distribution over the potential target shapes and assess the merit of a recently proposed approach of moving away from the threshold-based, binary definition of the CTV.

This approach, which accounts for CTV delineation uncertainties explicitly in planning, is to use what is known as the \textit{clinical target distribution} (CTD). The CTD is a distribution over the potentially tumorous voxels that for each voxel specifies a probability of tumor presence. In treatment planning, one may then use the CTD as voxel-wise weights in the optimization functions, to give higher priority to high-risk voxels. This approach was proposed by Shusharina et al. \cite{shusharina_clinical_2018} and has since then been explored further by Ferjancic et al. \cite{ferjancic_probabilistic_2021} and applied in a robust optimization context by Buti et al. \cite{buti_introducing_2021}. The idea of including voxel-wise probabilities of ROI occupation in optimization was proposed by Baum et al. \cite{baum_robust_2006} for managing overlapping margins in prostate treatments. Unkelbach and Oelfke then demonstrated that the approach was equivalent to minimizing the expected value of  certain objective functions with respect to the ROI-delineation uncertainty \cite{unkelbach_relating_2006}. Ideally, CTD-weighted optimization will assign dose even to low-probability regions if there is little conflict with sparing \textit{organs at risk} (OARs), while balancing OAR sparing and target coverage based on the probabilities in regions where the objectives conflict. Compared to more advanced approaches, e.g. the tumor control probability maximization by Bortfeld et al. \cite{bortfeld_probabilistic_2021}, this method has the advantage that the scaling of voxel weights in optimization preserves convexity and does not introduce any additional computational complexity.

In the present paper, we investigate the implications of CTD optimization compared to using some optimally chosen margin with respect to the underlying tumor infiltration model, the dose deliverability conditions, and the evaluation criteria. The comparison is based on the trade-offs between a target coverage criterion and penalties associated with dose to healthy tissue. The primary target coverage criterion considered is formulated as \textit{the probability of (almost) all parts of the target receiving (almost) the prescribed dose}. We have suspected that the low voxel-weight toward the edge of the CTD would result in plans with sub-optimal trade-offs between this criterion and the conflicting objectives, and that there rather exists some margin which is more efficient in the described sense. For cases when a proximal, critical OAR does not allow satisfactory values of the primary target coverage criterion, we consider additional criteria. In addition, we view the methods in the light of previously developed frameworks for managing geometric uncertainties, to better understand and compare the methods.

\section{Method} \label{method_section}

\subsection{Notation}

Any treatment planning problem in the present paper is treated as an optimization problem based on a discretization of the patient geometry into a grid of $m$ voxels. The set of ROIs, denoted by $R$, is partitioned into the set of OARs and the set of targets, denoted by $O$ and $T$, respectively. The voxel index set of any ROI $r$, $r \in R$, is then denoted by $\mathcal{R}_r$ or more specifically by $\mathcal{O}_r$ or $\mathcal{T}_r$ if the type of the ROI is known. Any voxel index $i \in \mathcal{R}_r$ has a corresponding \textit{relative volume} $v_{r,i}$ which is the ratio between the volume of voxel $i$ that belongs to $r$, and the volume of $r$. It follows that $\sum_{i \in \mathcal{R}_r} v_{r,i} = 1$. For the purposes of the present paper, it is useful to also define the absolute volumes $\Tilde{v}$ for which it holds that $v_{r,i} = \Tilde{v}_{r,i}/V_r, \forall r \in R, i \in \mathcal{R}_r,$ where $V_r$ is the volume of $r$.

\subsection{Preliminaries}

The \textit{dose-at-volume} (DaV) will be used to evaluate target coverage. For an ROI $r$, it is defined as the greatest dose that is received by at least the fraction $v$ of its volume:

\begin{equation}
    \text{DaV}^r_v(d) = \max \{ d' \in \mathbb{R} : \sum_{i \in \mathcal{R} : d_i \geq d'} v_{r,i} \geq v \},
\end{equation}

where the dose vector $d \in \mathbb{R}^m$ is a function of the optimization variables $x \in \mathcal{X}$ and $\mathcal{X} \subset \mathbb{R}^n$ is the feasible set, dependent on the delivery technique. DaV is non-convex and inherently hard to optimize, and will in this work thus only be used as a target coverage evaluation criterion. Instead, the optimization is performed with respect to presumably correlated surrogate functions. In the present paper, the standard notation $\text{D}_{100v}$ is used for $\text{DaV}_v$ when presenting results. To limit dose to some OAR $r$, we employ maximum dose functions of the form

\begin{equation}
    f_{r} \coloneqq \sum_{i \in \mathcal{O}_r} \Tilde{v}_{r,i} (d_i - \hat{d}_r)_+^2,
    \label{max_dose}
\end{equation}

where $\hat{d}_r$ denotes a reference dose level ideally not to be exceeded and $(\cdot)_+$ is the positive part operator. Function \eqref{max_dose} is convex and thus suitable for optimization. Since the focus in this work is on implications on target coverage, maximum dose functions are not only used in optimization but also for evaluation of over dosage. In what follows, analogous notation is used for dose-promoting functions in target volumes.

\subsection{Problem Setting}

\begin{figure}
\centering
\subfloat[1D]{%
  \begin{tikzpicture}[scale=0.8]
\begin{axis}[
    axis lines = middle,
    xmin = 0, xmax = 20, ymin = 0, ymax = 120,
    xlabel = \(z\),
    ylabel = {\(d\)},
    xtick distance = 6,
    clip=false,
]

\draw [decorate, decoration={brace,raise=0.6cm}] (6,0) -- (0,0) node [midway,yshift=-0.9cm] {GTV};
\draw [decorate, decoration={brace,raise=1.2cm}] (12,0) -- (0,0) node [midway,yshift=-1.5cm] {CTD};
\draw [decorate, decoration={brace,raise=0.6cm}] (18,0) -- (12,0) node [midway,yshift=-0.9cm] {OAR};

\draw [dashed, blue] (0, 100) -- (9, 100);
\draw [dashed, blue] (9, 100) -- (9, 0);
\draw [dashed, blue] (9, 0) -- (18, 0);
\draw [very thick, red] (0,0) -- (6,0);
\draw [very thick, orange] (6,0) -- (12,0);
\draw [very thick, blue] (12,0) -- (18,0);
\draw [thick, stealth-stealth] (7, 50) -- (11,50);

\end{axis}
\end{tikzpicture}
\label{1d_phantom}
}
\subfloat[3D]{%
  \begin{tikzpicture}[scale=0.8]
    \filldraw [color=orange, fill=orange!50, very thick] (0,0) circle (80pt);
    \filldraw [color=red, fill=red!50, very thick] (0,0) circle (40pt);
    \filldraw [color=blue, fill=blue!50, very thick] (0,88pt) circle (40pt);
    \node at (0, 0) {GTV};
    \node at (0, 88pt) {OAR};
    \node at (0, -60pt) {CTD};
    \end{tikzpicture}
\label{3d_phantom}
}
\caption{The phantom geometries under consideration.}
\end{figure}
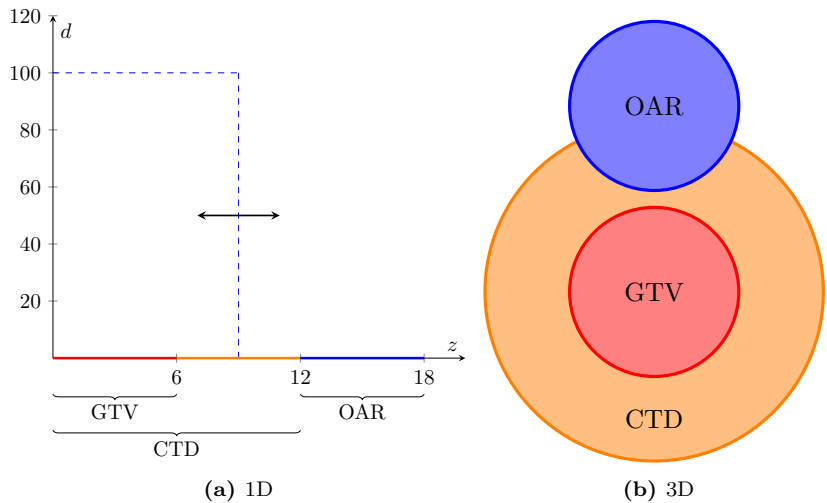

We consider patient phantoms with a known GTV and OAR. The CTD extends away from the edge of the GTV and toward the OAR, and is the region at risk of microscopic tumor infiltration. The External ROI is defined to comprise the full patient volume. Both a 1D and a 3D setup are considered to first display the basic properties of the methods in the simplest setting, and then show the implications in a setting which is more realistic. The 1D setup and a 2D slice of the 3D setup are illustrated in Figures \ref{1d_phantom} and \ref{3d_phantom}, respectively. To avoid ambiguity when assessing resulting plans, we do not account for overlap between the conflicting regions. Instead, we assume that the OAR acts as an impenetrable barrier for the tumor infiltration, and should ideally not receive any dose.

\subsubsection{Tumor Expansion Model}

To model microscopic tumor spread we consider an isotropic infiltration model away from the GTV, proposed in \cite{buti_introducing_2021}. Therein, the tumor spread can be characterized by a single random variable $S$ with probability density function $\rho(s)$. The probability of the tumor to have spread beyond a certain distance $s$ from the GTV is then given by

\begin{equation}
    \mathbb{P} (S \geq s) = \int_s^\infty \rho(x) dx.
\label{cdf}
\end{equation}

\subsubsection{Discretization into Scenarios} \label{scenarios}

For computational purposes, the distribution of $S$ is discretized into a scenario set $\mathcal{S}$. For distributions with negligible tail, this discretization involves setting a cut-off value to be considered as the worst case, and renormalizing the scenario probabilities, which are then denoted by $q_s, \forall s \in \mathcal{S}$. In the following, each scenario $s$ maps to a voxel index set $\mathcal{T}_s$ and the corresponding relative volumes. In 3D, this mapping is defined by the ROI algebra functionality in RayStation (RaySearch Laboratories AB, Stockholm).

\subsection{Establishing and Managing Conflicting Objectives} \label{MCO_formulation}

Given the unknown extent of the true target $t_S$, one must define appropriate criteria by which to evaluate a given treatment plan. In the present paper, we assume that any tumorous voxel should ideally receive almost the full prescribed dose, by considering the target coverage metric

\begin{equation}
    f^{v,\gamma}_{t_S}(d(x)) \coloneqq \mathbb{P}(\text{DaV}^{t_S}_v(d(x)) \geq \gamma \hat{d}),
    \label{target_coverage_probability}
\end{equation}

which is the probability that the DaV is at least $\gamma \hat{d}, \gamma \in [0, 1]$. Since our concern is the minimum or near-minimum dose, we typically set both $\gamma$ and $v$ close to or equal to $1$.

There are typically a number of criteria of minimal dose to healthy tissue which stand in conflict with maximizing the target coverage criterion \eqref{target_coverage_probability}. Thus, we add an objective that is a weighted sum of high dose penalties, resulting in the bi-objective formulation

\begin{mini}|l|
{x \in \mathcal{X}}{(-f^{v,\gamma}_{t_S}(d(x)), \sum_{r \in O} w_r f_r(d(x)))}{}{},
\label{MCO}
\end{mini}

where each $f_r$ is a maximum dose function with some associated non-negative importance weight $w_r$. Because DaV is intractable in optimization, we consider instead a special case of problem \eqref{MCO} with $\gamma = v = 1$ (the probability that the minimum target dose is at least $\hat{d}$).

\begin{mini}|l|
{x \in \mathcal{X}}{(-f^{1, 1}_{t_S}(d(x)), \sum_{r \in O} w_r f_r(d(x)))}{}{}.
\label{special_case_MCO}
\end{mini}

Problem \eqref{special_case_MCO} is a \textit{multi-criteria optimization} (MCO) problem with a discontinuous objective that could be modelled with $|\mathcal{S}|$ binary variables, where each would be used to indicate target coverage in a scenario. Although problem \eqref{special_case_MCO} could therefore be solved by using a weighted sum of the two objectives, doing so reduces the control over the final value of $f^{1, 1}_{t_S}$. While we shall return to a weighted sum approach in what follows, we now use the $\epsilon$-constraint method for MCO problems, as described in Appendix \ref{epsilon-constraint method}, for generating a set of Pareto-optimal solutions to problem \eqref{special_case_MCO}. Thus, for $J$ clinically acceptable values of $\epsilon_j \in [0, 1], j = 1, \dots, J$, we solve

\begin{mini}|l|
{x \in \mathcal{X}}{\sum_{r \in O} w_r f_r(d(x))}{}{}
\addConstraint{\mathbb{P}(d_i(x) \geq \hat{d}, \forall i \in \mathcal{T}_S) \geq \epsilon_j.}
\label{epsilon_formulation}
\end{mini}

The simplicity of the isotropic infiltration model means that solving problem \eqref{epsilon_formulation} for a given target coverage probability $\epsilon_j$ is not difficult. Even though $\mathcal{T}_S$ is modelled as random, the set of voxels $\mathcal{T}_{\text{CTV}_{\epsilon_j}}$ that then require dose is cheaply computed in advance. This can be seen from the fact that voxel sets corresponding to smaller values of $\epsilon_j$ are contained in those corresponding to larger values, or more formally: For any $i, j$ such that $\epsilon_i \leq \epsilon_j$, it follows that $\mathcal{T}_{\text{CTV}_{\epsilon_i}} \subseteq \mathcal{T}_{\text{CTV}_{\epsilon_j}}$. Thus, problem \eqref{epsilon_formulation} may equivalently be written as the deterministic problem

\begin{mini!}|l|
{x \in \mathcal{X}}{\sum_{r \in O} w_r f_r(d(x))}{\label{equivalent_epsilon_formulation}}{}
\addConstraint{d_i(x)}{\geq \hat{d}, \forall i \in \mathcal{T}_{\text{CTV}_{\epsilon_j}}.\label{linear_constraints}}
\end{mini!}

For computational purposes, we replace the linear inequalities \eqref{linear_constraints} by a single function inequality:

\begin{equation}
    f_{\text{CTV}_{\epsilon_j}}(d(x)) \coloneqq \sum_{i \in \mathcal{T}_{\text{CTV}_{\epsilon_j}}} \Tilde{v}_{\text{CTV}_{\epsilon_j},i} \cdot (\hat{d} - d_i(x))_+^2 \leq 0.
    \label{quadratic_underdosage_constraint}
\end{equation}

In summary, we have replaced the true problem \eqref{MCO} with a surrogate problem \eqref{equivalent_epsilon_formulation}, in return for tractability. The $\epsilon$-constraint methods ensures that the set of solutions to problem \eqref{equivalent_epsilon_formulation} belong to the Pareto set of problem \eqref{special_case_MCO}. The set of associated solutions will act as a reference for the methods presented next.

\subsection{Enumeration of CTV Margins} \label{enumeration}

Although the $\epsilon$-constraint method can be used to bound the target coverage probability in our simple model, it may not be most desirable to do so. First, it is possible that maximizing function \eqref{target_coverage_probability} with $v = \gamma = 1$ requires too much dose, and that striving for a more lenient, clinically relevant goal (e.g. $\text{D}_{98} > 0.95 \hat{d}$) could reduce integral dose and overdosage to OARs. Second, our empirical experience shows that finding a feasible solution to a problem with such a non-linear constraint (or equivalently a large number of linear constraints) may be computationally inefficient in practice.

Instead, we solve a relaxation of problem \eqref{equivalent_epsilon_formulation} by moving the left hand side of \eqref{quadratic_underdosage_constraint}, scaled by an importance weight $w_t > 0$, to the objective. With this relaxation we arrive at conventional margin-based planning where the CTV margin is determined by the probability level $\epsilon_j$.

\begin{mini}|l|
{x \in \mathcal{X}}{w_t f_{\text{CTV}_{\epsilon_j}}(d(x)) + \sum_{r \in O} w_r f_r(d(x))}{\label{enumeration_formulation}}{}
\end{mini}

\subsection{The CTD as Voxel-Wise Weights in the Objective} \label{ctd_weights}

We now arrive at the CTD approach which consists of a formulation which is similar to problem \eqref{enumeration_formulation} but with the first term in the objective replaced by $ w_t f_{\text{CTD}}(d(x))$, where

\begin{equation}
   f_{\text{CTD}}(d(x)) \coloneqq \sum_{i \in \mathcal{T}_{\text{CTD}}} p_i \Tilde{v}_{\text{CTD},i} \cdot (\hat{d} - d_i(x))_+^2,
    \label{ctd-objective}
\end{equation}

in which each $p_i, \forall i \in \mathcal{T}_{\text{CTD}}$ denotes the probability of voxel $i$ being tumorous. Each $p_i$ is calculated from equation \eqref{cdf} using voxel-wise distances to the surface of the GTV.

\subsection{The Methods in a Robust Optimization Context}

Using the scenario discretization in Section \ref{scenarios}, the tumor infiltration problem can be seen through the robust optimization framework developed for geometric uncertainties by Fredriksson \cite{fredriksson_characterization_2012}. In this framework, margin-based planning is equivalent to planning with respect to a single scenario, or to worst-case optimization if the tail of the probability distribution which falls outside the margin is negligible. Since a larger margin represents a scenario with a target voxel index set that contains those of smaller margins, the largest considered margin will always make up the worst case with respect to minimum dose functions. Varying the size of the margin is then equivalent to varying the length of the neglected distribution tail. Then, a discretization made so that the largest target expansion scenario maps to $\text{CTV}_{\epsilon_{J}}$ implies that

\begin{equation}
    f_{\text{CTV}_{\epsilon_J}}(d(x)) = \underset{s \in \mathcal{S}}{\max} \{ f_{t_S}(d(x)) \}. 
\end{equation}

In contrast, using the CTD objective is analogous to the target presence probability approach in Baum et al. \cite{baum_robust_2006}. Similarly to what was noted by Unkelbach and Oelfke \cite{unkelbach_relating_2006}, the CTD objective \eqref{ctd-objective} is approximately equal to the expected value of the minimum dose function over the scenario set, making the CTD approach analogous also to the expected-value minimization in Fredriksson \cite{fredriksson_characterization_2012}.

\begin{equation}
    f^{\text{CTD}}(d(x)) \approx \mathbb{E}[\sum_{i \in \mathcal{T}_{S}} \Tilde{v}_{t_S,i} \cdot (\hat{d} - d_i(x))_+^2] = \sum_{s \in \mathcal{S}} q_s \sum_{i \in \mathcal{T}_s} \Tilde{v}_{t_s,i} \cdot (\hat{d} - d_i(x))_+^2.
\end{equation}

The approximation is increasingly accurate with the number of voxels $m$ and the number of scenarios $|\mathcal{S}|$.

\section{Results} \label{results_section}

\subsection{Patient Cases}

To compare the optimization methods outlined in Section \ref{method_section}, each method has been implemented on a 1D and a 3D phantom, respectively. For the 1D case, a phantom was modeled as the interval $[0, L]$, like in Figure \ref{1d_phantom}, with $L = 18$ cm. The GTV and an OAR were represented as the sub-intervals $[0, L/3]$ and $[2L/3, L]$, respectively. The CTD was characterized as an expansion of the GTV by $S$ cm, with $S$ uniformly distributed in the interval $[0, L/3]$. Dose delivery was modelled with equally spaced Gaussian shaped spots with standard deviation $\sigma = 0.5$ cm and inter spot distance $1.5 \sigma$. The leftmost spot was centered at $l = -1$. For each approach, the model was discretized into $m = 9000$ voxels using MATLAB (2021a, The MathWorks, Inc., Natick, Massachusetts, United States), and solved with the \textit{sequential quadratic programming} (SQP) solver SNOPT (7.7 Stanford Business Software, Stanford, California).

In 3D, a model of a water phantom comprising $m = 168 \times 177 \times 177$ voxels was created in a research version of the treatment planning system RayStation. The GTV and an OAR were modelled as two spheres, both of radius 1.5 cm and 0.3 cm apart. The CTD was then characterized by an expansion of the GTV by $S$ from a truncated normal distribution between 0 and 2 cm, with mean 0.6 cm and standard deviation 0.4 cm, followed by a subtraction of the overlap with the OAR. The phantom setup is shown in Figure \ref{3d_phantom}. To limit bias toward a specific modality, results were generated for both \textit{intensity-modulated proton therapy} (IMPT), with two laterally opposed beams, and tomotherapy. The problems were solved using 200 iterations of RayStation's internal SQP algorithm, which was sufficient for convergence to a resulting dose distribution. For tomotherapy, the resulting optimization problems were non-convex due to a filtering step: Leaf opening times were either set to 0 or bounded from below at iteration 100, depending on their current value. Spot weight filtering was not performed for IMPT, resulting in convex problems.

\begin{table}[!ht]
\centering
\begin{tabular}{ *5c }
Case & $r$ & $w_r$ & \multicolumn{2}{c}{$\hat{d}_r$} \\ \toprule
& & & 1D & 3D \\ \midrule

 & External & 100 & 100 & 100 \\
A & External & 0.01 & 0 & 0 \\ 
 & OAR & 1 & 0 & 0 \\ \midrule
 & External & $\infty$ & 101 & 105 \\
B & External & 0.01 & 0 & 0 \\ 
 & OAR & $\infty$ & 50 & 75 \\
 & OAR & 1 & 0 & 0 \\ \bottomrule
\end{tabular}
\caption{Parameter values of the dose-limiting maximum dose functions.}
\label{oar_functions}
\end{table}

In Section \ref{MCO_formulation} we considered a special case of \eqref{target_coverage_probability} in which $\gamma$ and $v$ are both close to 1. In this case, the functions used to penalize high dose and their corresponding weights were chosen as in Case A in Table \ref{oar_functions}. When evaluating target coverage, the metrics of interest were $\mathbb{P} (\text{D}_{100} > 0.99 \hat{d})$ in 1D and $\mathbb{P} (\text{D}_{98} > \gamma \hat{d})$ for $\gamma = 0.95, 0.97, 0.99$ in 3D. To address cases when such strict criteria may be physically incompatible with strict limitations on dose to the OAR, the implications of parameter values chosen as in Case B were also investigated. There, a maximum dose constraint on the OAR was used to bound the OAR dose from above. In both cases, the target metrics for the resulting plans were evaluated by considering 1000 samples of $S$, for each of which $\text{DaV}$ values were calculated. The currently considered target metric was then compared to the weighted sum of maximum dose functions used as dose-limiting objectives, normalized by the maximum across the plans for the current case.

\subsection{1D Results}

\begin{figure}[!ht]
\centering
\subfloat[Case A: Trade-offs.]{%
\includegraphics[clip,width=0.5\textwidth]{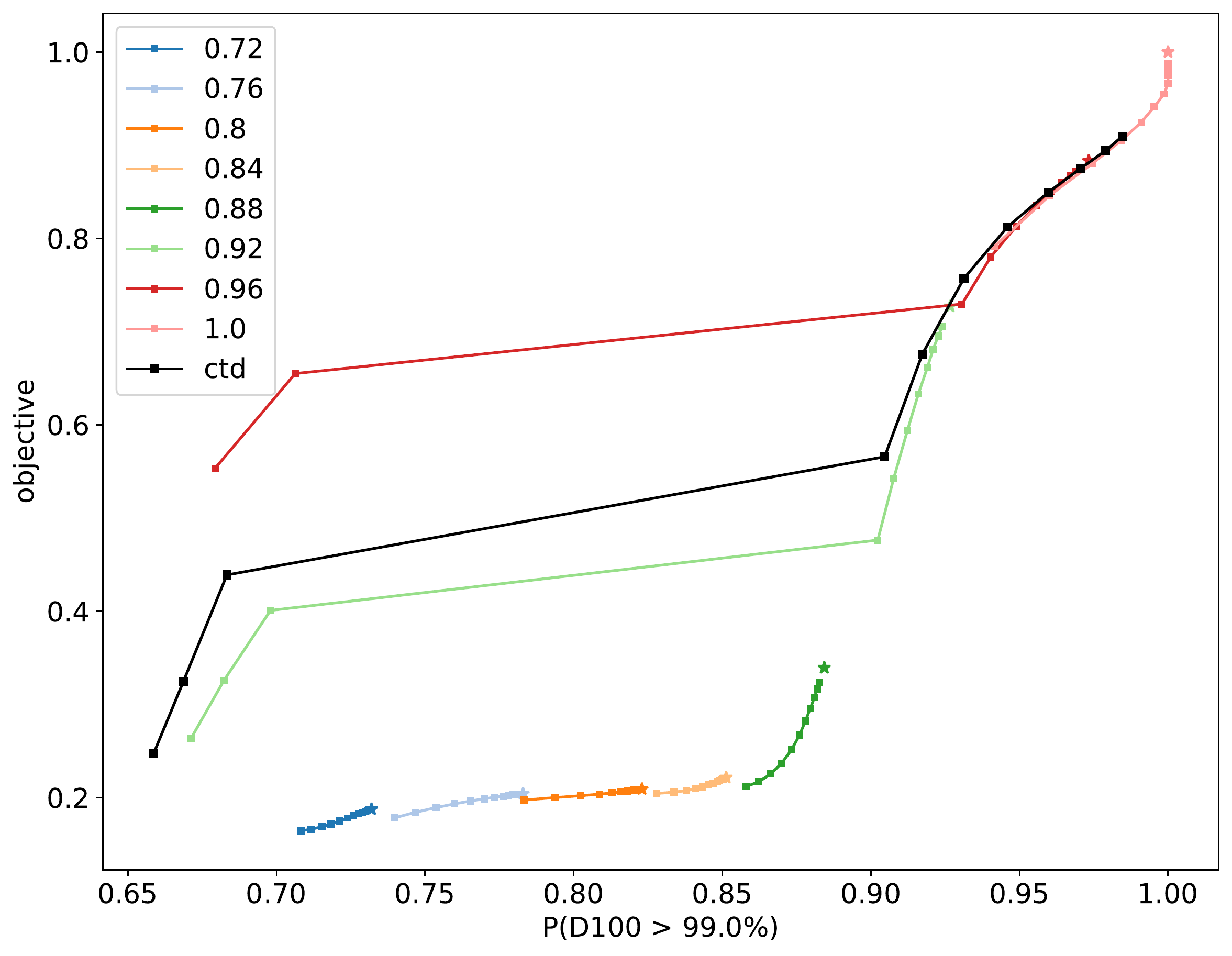}%
  \label{1d_trade-off}
}
\subfloat[Case A: $w_t = 1600$.]{%
\includegraphics[clip,width=0.5\textwidth]{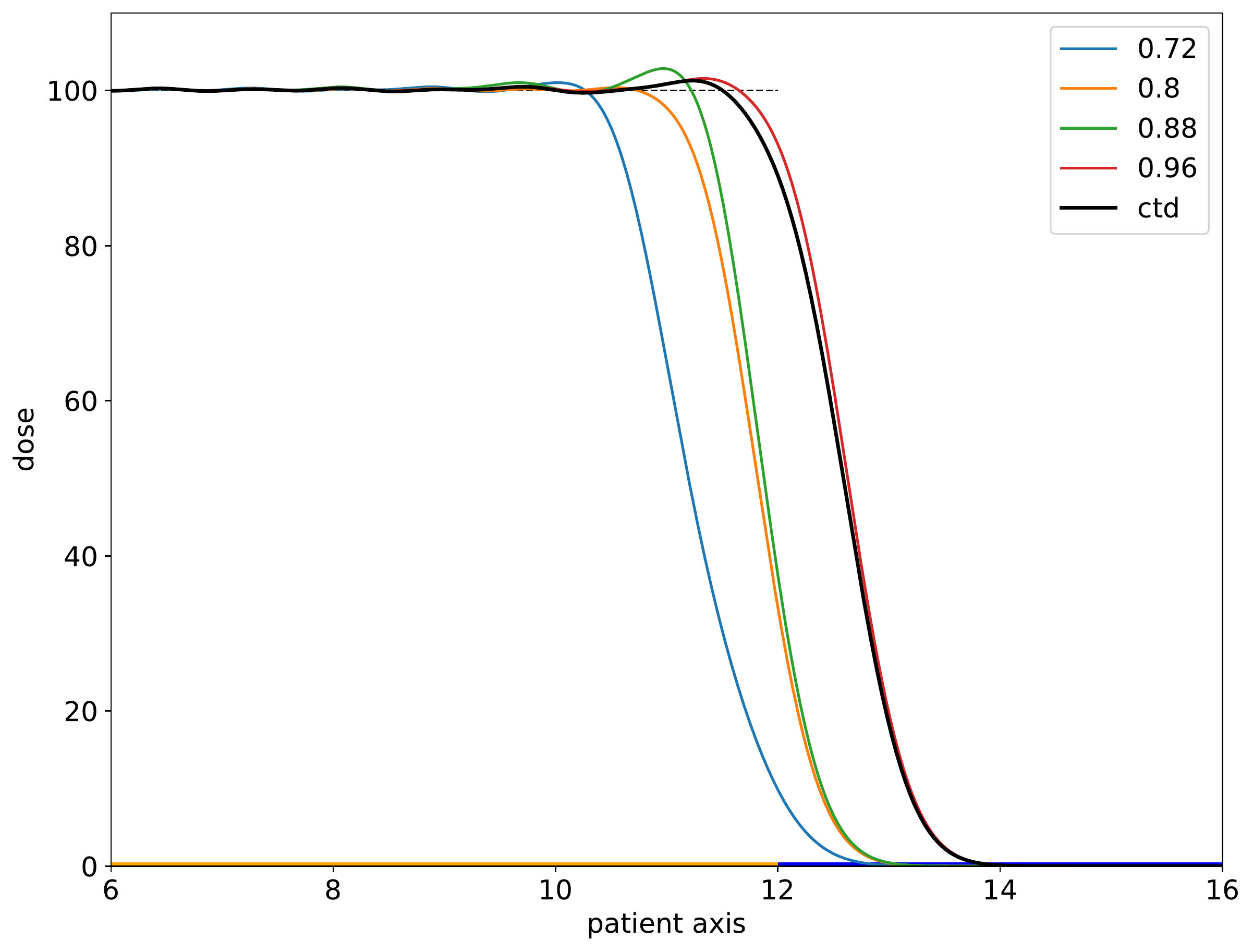}%
  \label{1d_results}
 }
 
\subfloat[Case A: $w_t = 50$.]{%
\includegraphics[clip,width=0.5\textwidth]{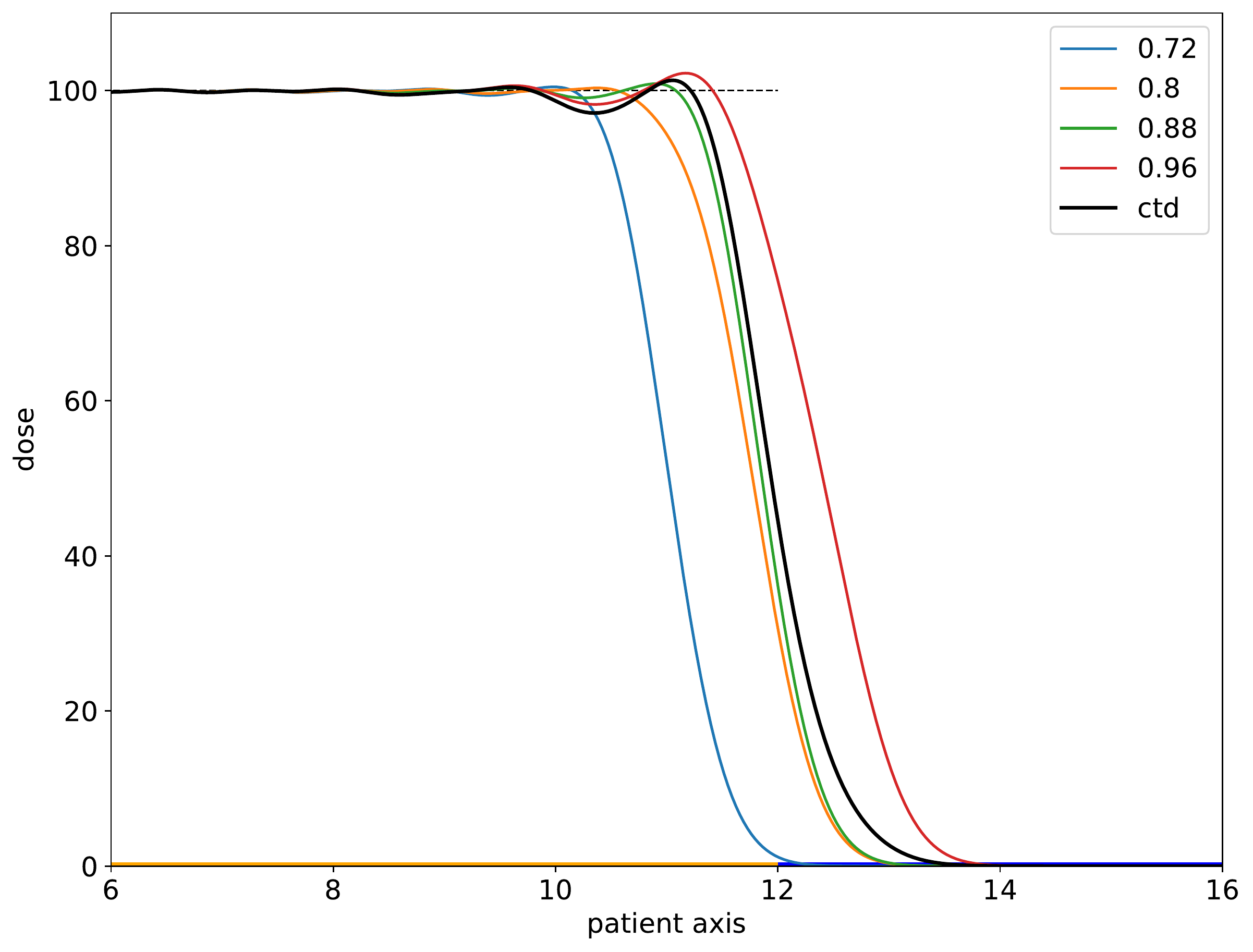}%
  \label{early_fall-off}
}
\subfloat[Case B: $w_t = 100$.]{%
\includegraphics[clip,width=0.5\textwidth]{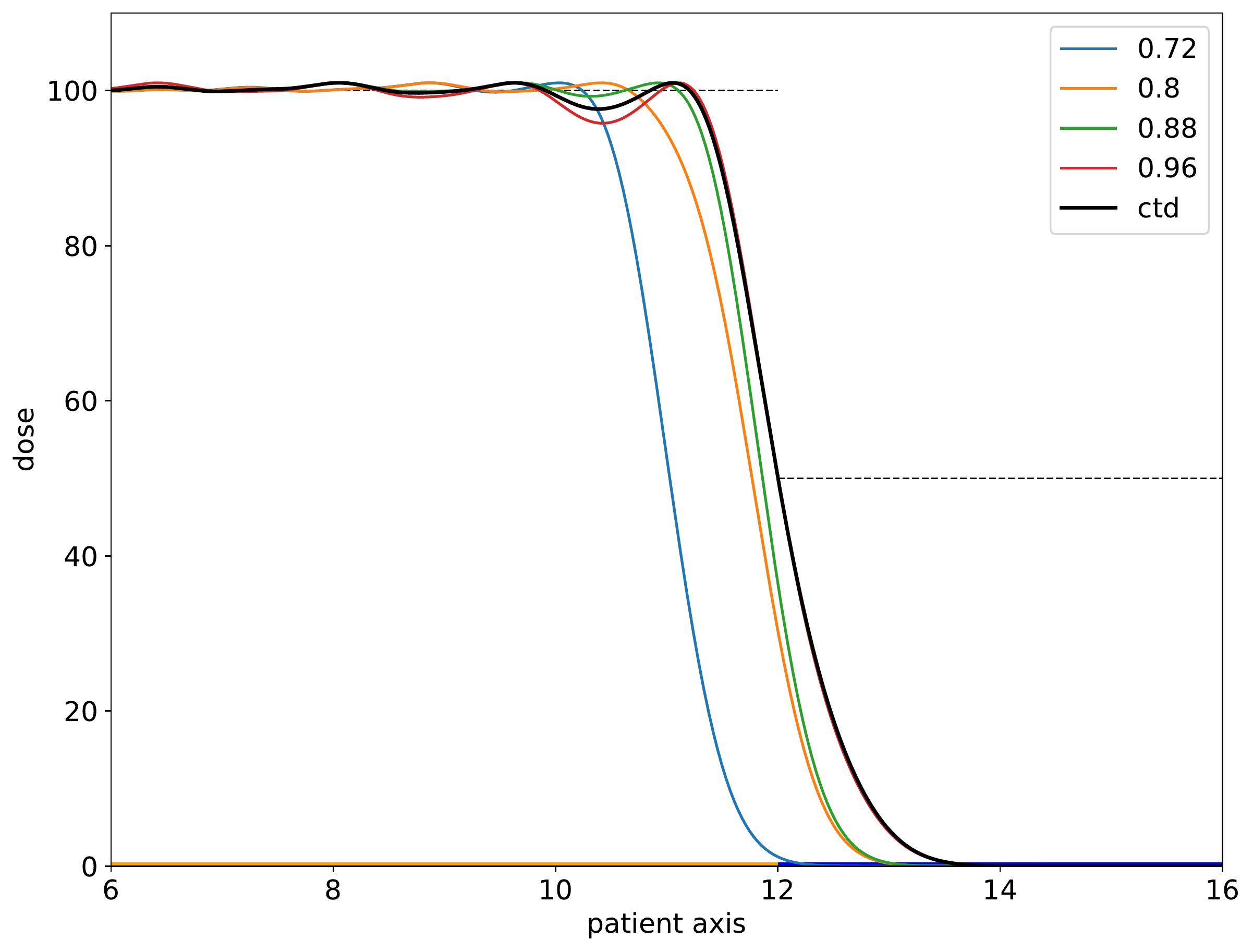}%
  \label{1d_results_strict}
 }
\caption{1D: (a) Trade-offs between the target coverage metric and the normalized
dose-limiting objective for each of the methods with $w_t \in [50, 51200]$, for Case A. Plans using the constraints from problem \eqref{equivalent_epsilon_formulation} are marked with a star. (b) Corresponding doses for the plans with $w_t = 1600$. (c) Corresponding doses for the plans with $w_t = 100$. (d) Doses for each of the methods in Case B.}

\end{figure}

Figure \ref{1d_trade-off} shows the trade-off curves of the plans in the evaluation criteria space for a set of various target weights $w_t$. Notably, for high values of $w_t$, results are similar between the methods, in the sense that the trade-off curves are close in the criteria space. This is emphasized further by the similarity of the dose profiles between the CTD plan and the margin-based plans in Figure \ref{1d_results}. For low $w_t$, the doses from the CTD plan are clearly not Pareto optimal due to the early and slow dose fall-off, which can also be seen from the dose profiles in Figure \ref{early_fall-off}. Going forward, we mainly consider results for values of $w_t$ which are empirically determined to result in CTD plans with satisfactory target coverage, replicating the approach in Buti et al. \cite{buti_introducing_2021}. We believe that this is similar to what could be expected from a manual planner in a clinical planning process.

Figure \ref{1d_results_strict} shows the 1D dose profiles for Case B. Clearly, larger margins suffer more from variation in dose close to the conflict region, while smaller margins result in more homogeneous dose. The dose of the CTD plan is a compromise between the extremes, and better at maintaining high dose toward the center of the target compared to the doses from using large margins.

\subsection{Case A: Near Minimum Target Dose as First Priority}

\begin{figure}[!ht]
    \centering
    \includegraphics[width=\textwidth]{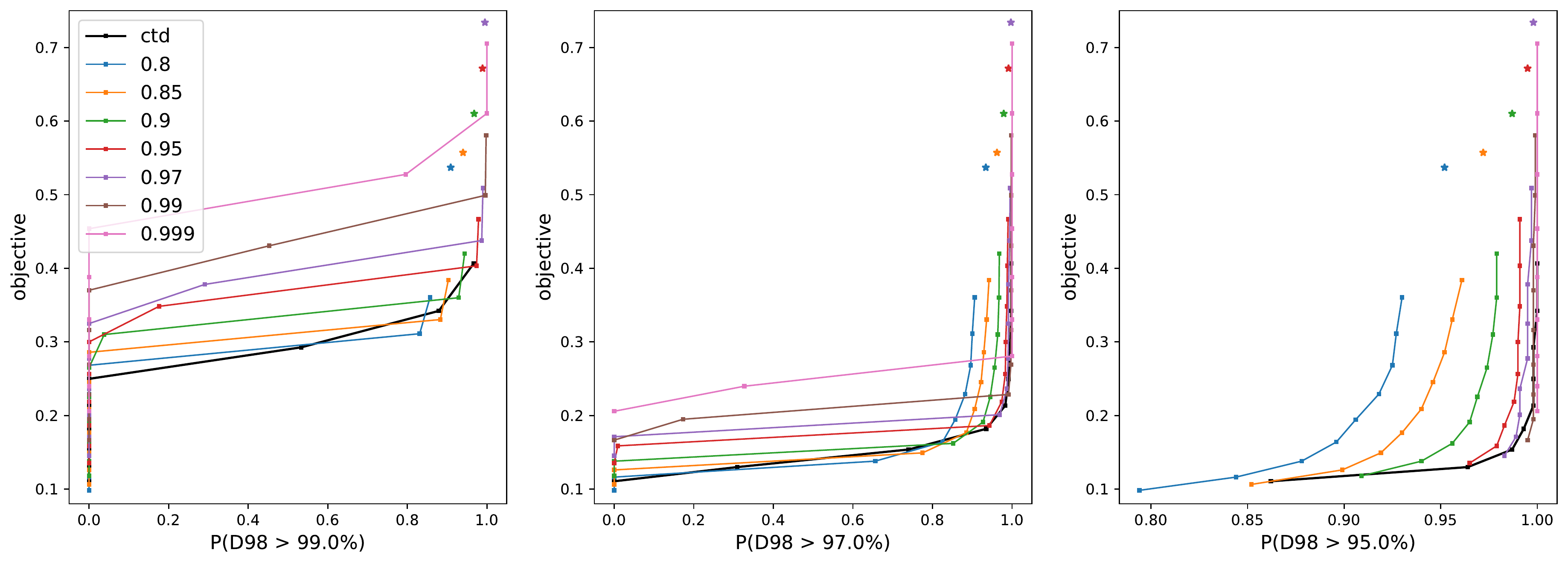}
    \caption{Case A: Trade-offs between target coverage metrics and the normalized dose-limiting objective for IMPT. $w_t \in [50, 12800]$. Plans using the constraints from problem \eqref{equivalent_epsilon_formulation} are marked with a star.}
    \label{caseA_proton_trade-offs}
\end{figure}

\begin{figure}[!ht]
    \centering
    \includegraphics[width=\textwidth]{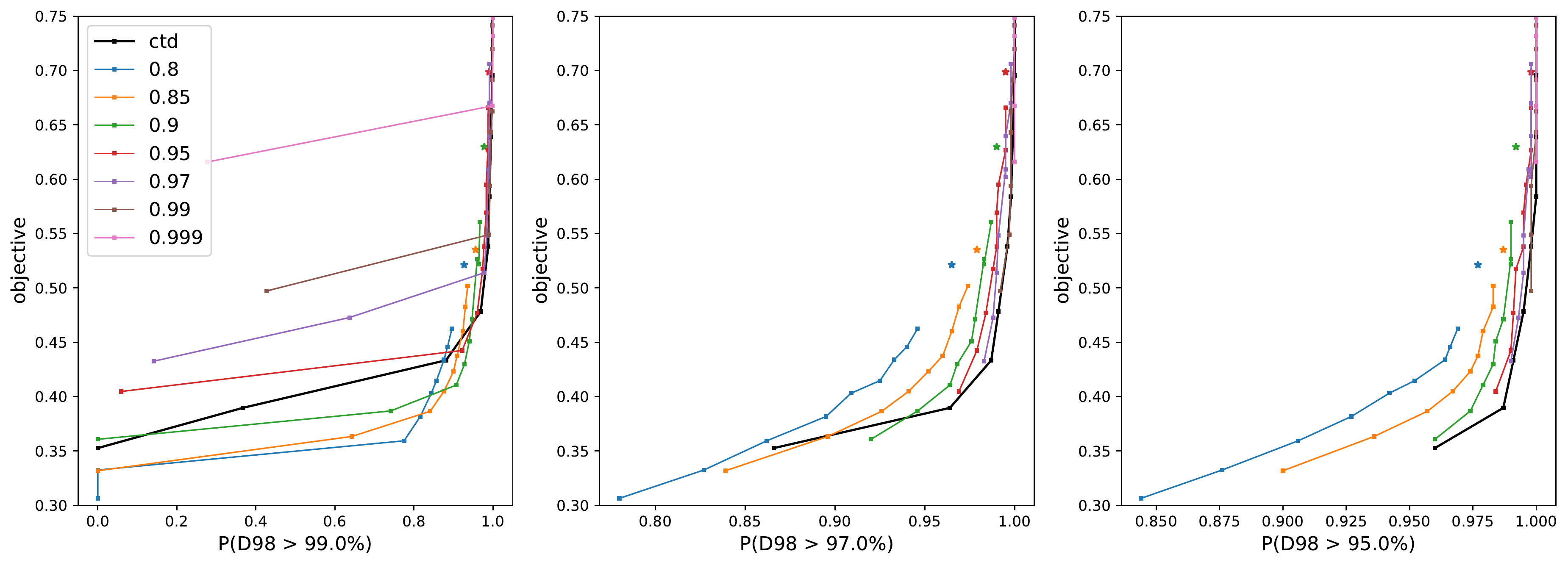}
    \caption{Case A: Trade-offs between target coverage metrics and the normalized dose-limiting objective for tomotherapy. $w_t \in [50, 12800]$. Plans using the constraints from problem \eqref{equivalent_epsilon_formulation} are marked with a star.}
    \label{caseA_tomo_trade-offs}
\end{figure}

Figures \ref{caseA_proton_trade-offs} and \ref{caseA_tomo_trade-offs} show the trade-offs of the plans from each of the methods considering a set of various target weights $w_t$, for IMPT and tomotherapy, respectively. For the highest dose threshold ($\gamma = 0.99$), the trade-off curve from the CTD approach is less efficient than the convex hull formed by the margin-based plans. The difference seems to decrease with increasing $w_t$. In addition, it can be noticed that the trade-off curves from the CTD approach are increasingly competitive for more lenient target coverage criteria ($\gamma = 0.97, 0.95$), sometimes even more efficient than the convex hull formed by the margin-based plans. Notably, the plans based on formulation \eqref{equivalent_epsilon_formulation} are always inefficient with respect to the presented evaluation criteria.

\begin{figure}[!ht]
    \centering
\subfloat[$\text{CTV}_{0.99}$ margin, $w_t = 200$.]{%
  \includegraphics[trim={0 0.1cm 0.1cm 0.1cm},clip,width=0.5\textwidth]{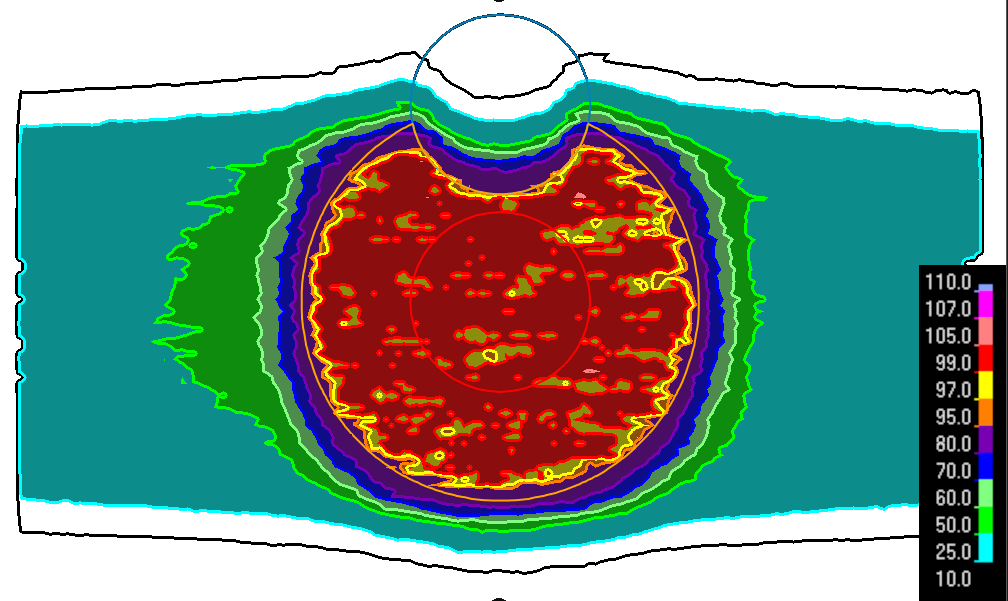}%
  \label{}
}
\subfloat[CTD, $w_t = 800$.]{%
  \includegraphics[trim={0 0.1cm 0.1cm 0.1cm},clip,width=0.5\textwidth]{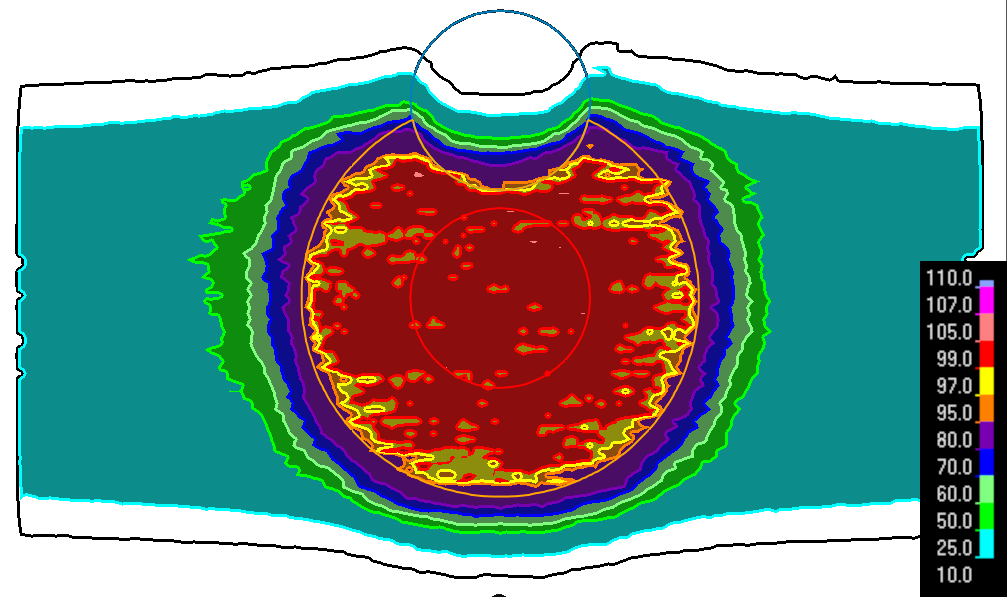}%
  \label{}
}

\subfloat[Dose difference, (b) - (a).]{%
  \includegraphics[clip,width=0.5\textwidth]{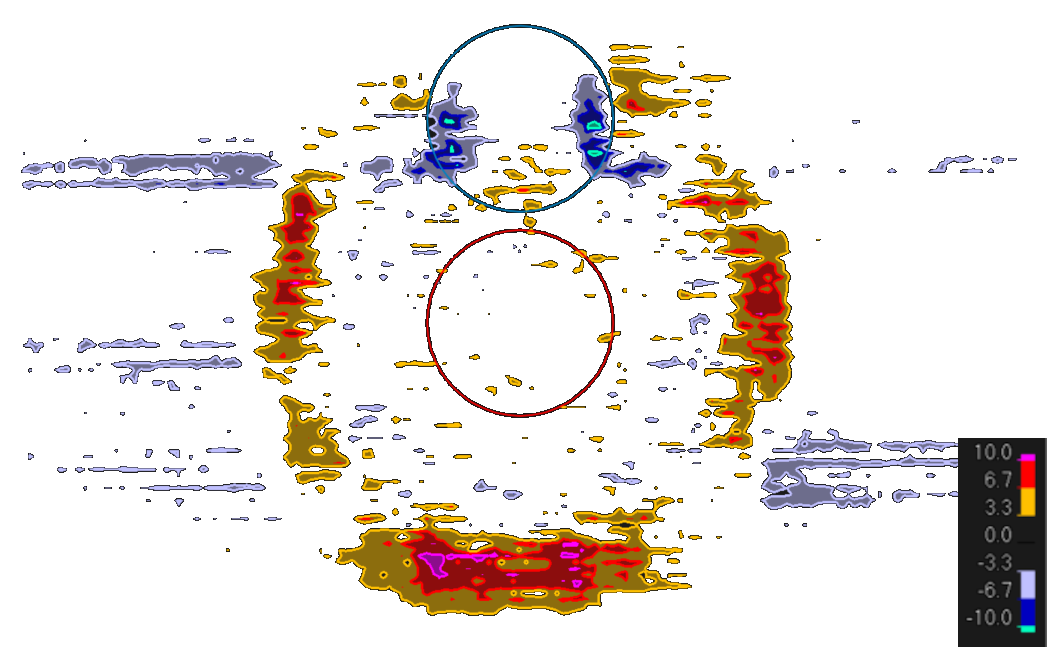}%
  \label{}
}
\subfloat[DVH curves for the External (green) and the OAR (blue). DVH bands between the 0.1- and 0.9-quantiles of the target sample distribution (red), as well as the DVH curve of the median (black).]{%
  \includegraphics[width=0.5\textwidth]{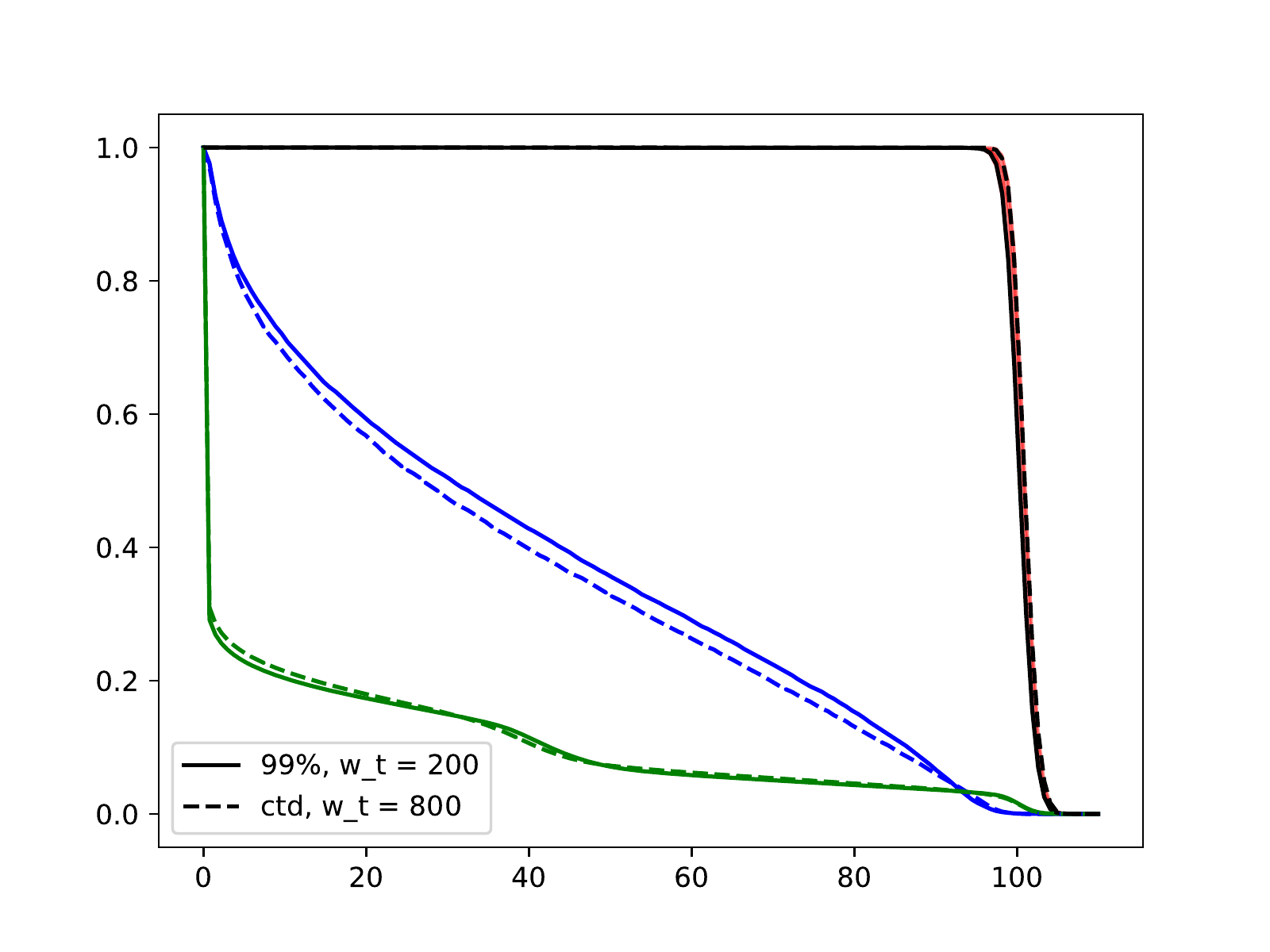}%
  \label{case_a_proton_DVHs}
 }
 \caption{Case A: Dose differences between the considered methods for IMPT.}
 \label{case_a_proton_doses}
\end{figure}

\begin{figure}[!ht]
    \centering
\subfloat[$\text{CTV}_{0.9}$ margin, $w_t = 200$.]{%
  \includegraphics[trim={0.1cm 0 0.1cm 0.2cm},clip,width=0.5\textwidth]{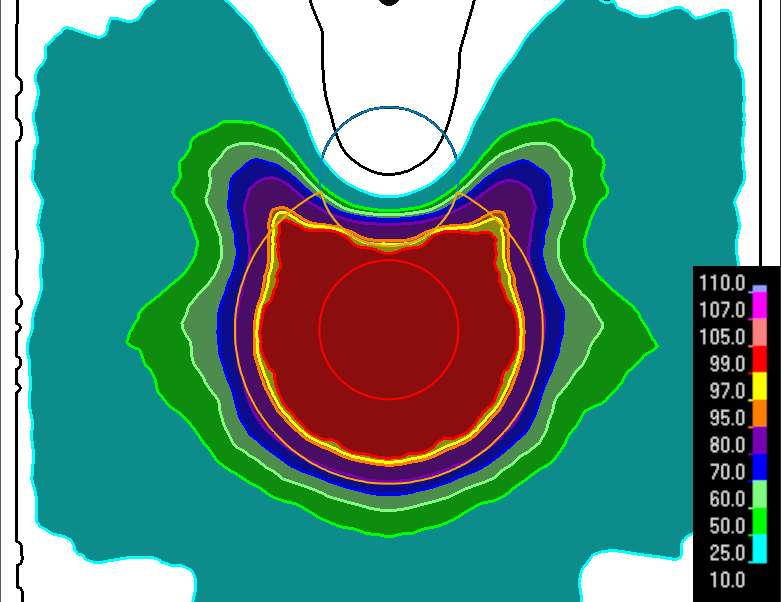}%
  \label{}
}
\subfloat[CTD, $w_t = 200$.]{%
  \includegraphics[trim={0.1cm 0 0.1cm 0.2cm},clip,width=0.5\textwidth]{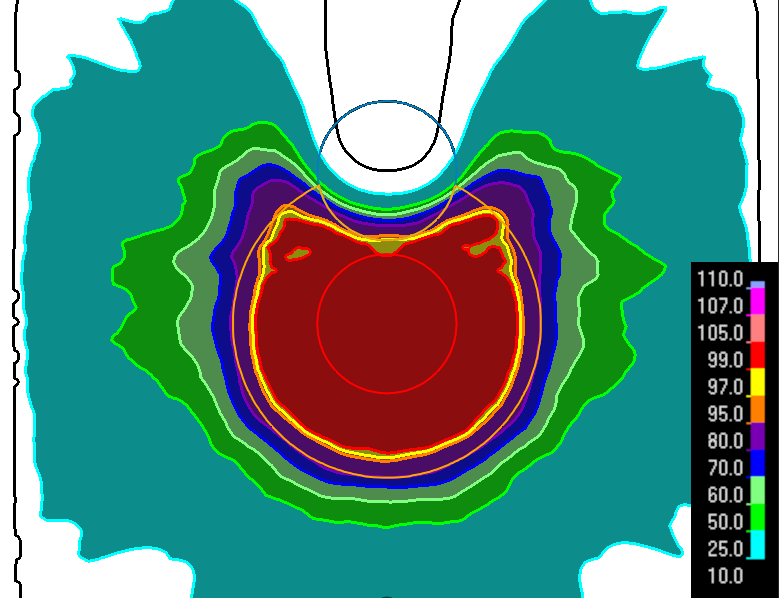}%
  \label{}
}

\subfloat[Dose difference, (b) - (a).]{%
  \includegraphics[clip,width=0.5\textwidth]{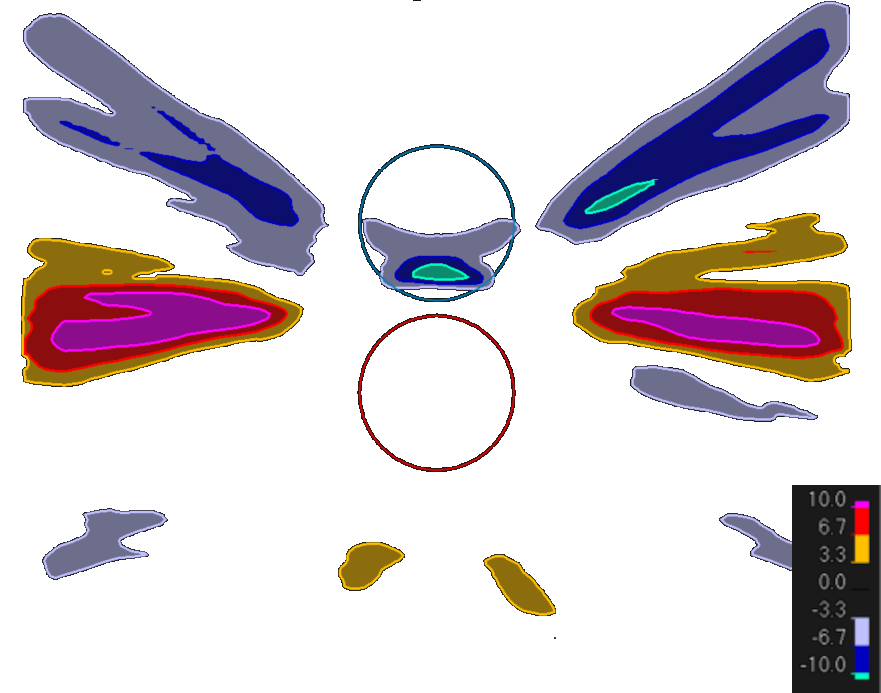}%
  \label{}
}
\subfloat[DVH curves for the External (green) and the OAR (blue). DVH bands between the 0.1- and 0.9-quantiles of the target sample distribution (red), as well as the DVH curve of the median (black).]{%
  \includegraphics[width=0.5\textwidth]{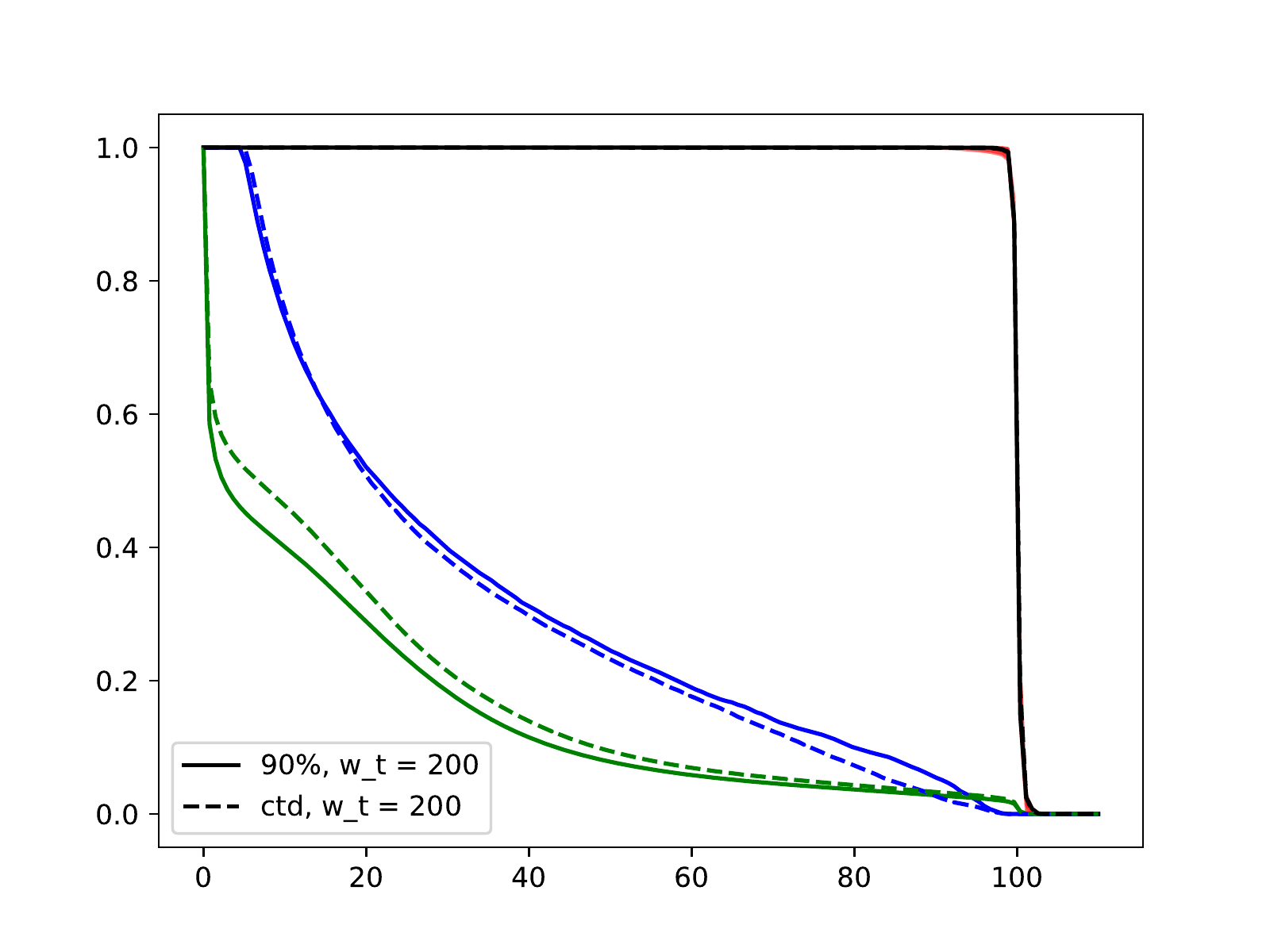}%
  \label{case_a_tomo_DVHs}
}
 \caption{Case A: Dose differences between the considered methods for tomotherapy.}
 \label{case_a_tomo_doses}
\end{figure}

Figures \ref{case_a_proton_doses} and \ref{case_a_tomo_doses} show example doses and corresponding \textit{dose-volume histograms} (DVHs) from plans for each of the methods for IMPT and tomotherapy, respectively. The plans have been chosen based on similar performance with respect to the trade-offs from Figures \ref{caseA_proton_trade-offs} and \ref{caseA_tomo_trade-offs}. For both modalities, the CTD-based plan achieves better sparing of the OAR for comparable target coverage. However, it seems to come at the cost of increased total dose, as seen especially from the DVH curve of the External for tomotherapy in Figure \ref{case_a_tomo_DVHs}. For IMPT, the CTD approach gives a noticeably higher dose near the edge of the CTD. For tomotherapy, the dose differences are larger for certain beam angles near the conflict region.

\newpage

\subsection{Case B: Strict OAR Constraint as First Priority}

\begin{figure}[!ht]
    \centering
    \includegraphics[width=\textwidth]{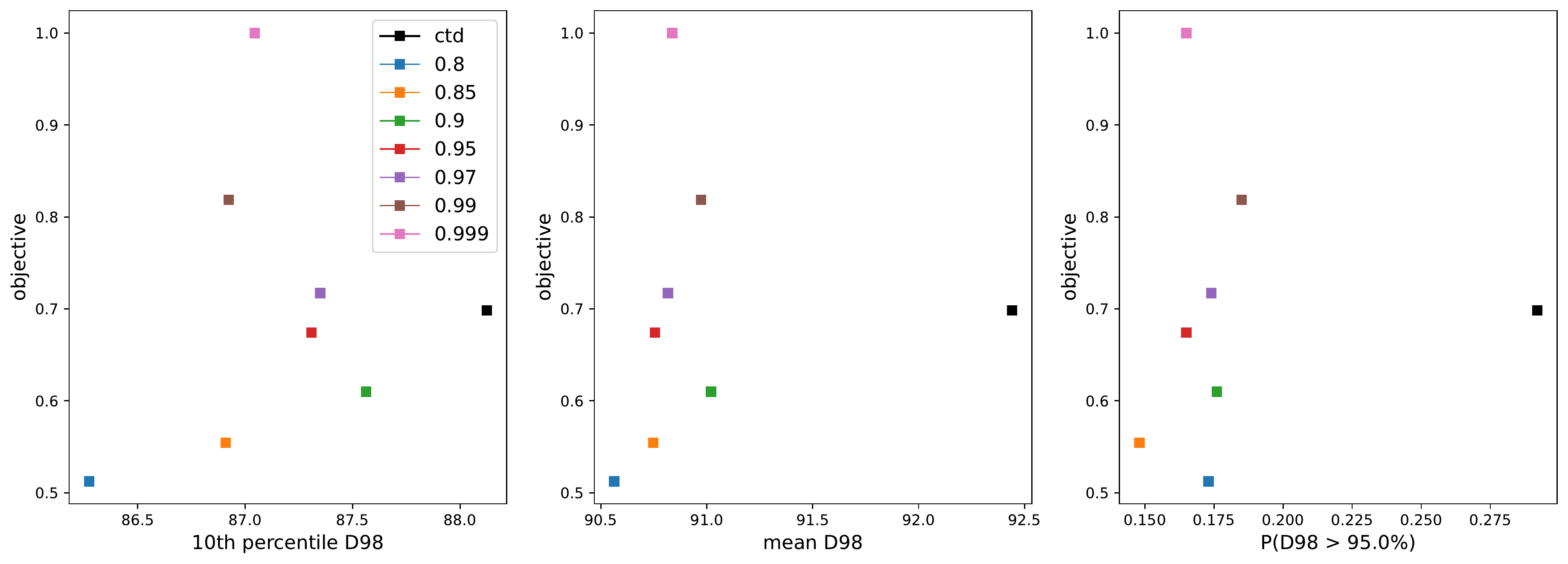}
    \caption{Trade-offs between target coverage metrics and the dose-limiting objective for IMPT.}
    \label{caseB_proton_trade-offs}
\end{figure}

\begin{figure}[!ht]
    \centering
    \includegraphics[width=\textwidth]{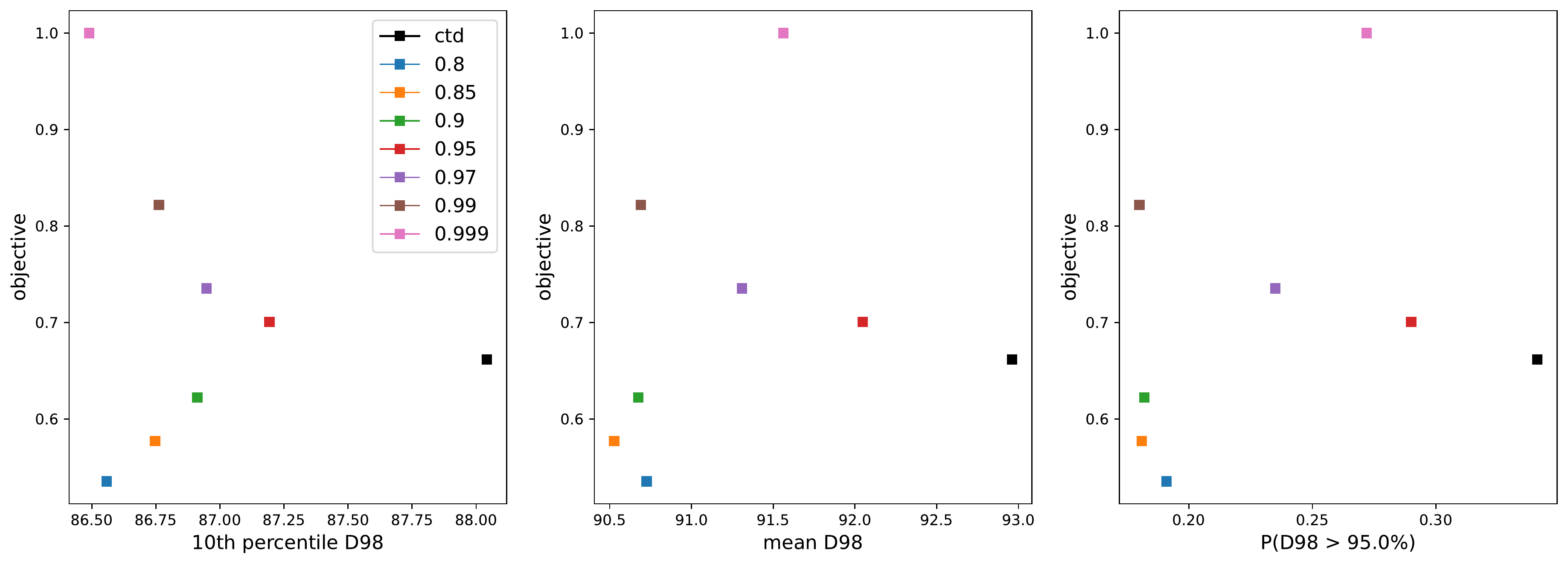}
    \caption{Trade-offs between target coverage metrics and the dose-limiting objective for tomotherapy.}
    \label{caseB_tomo_trade-offs}
\end{figure}

Given the constraint on maximum OAR dose, physical limitations on dose conformity make it physically impossible for the target coverage criterion \eqref{target_coverage_probability} to approach 1 given the values of $v$ and $\gamma$ in Case A. In addition, the probability of meeting a certain DaV threshold may be insufficient to score a plan in terms of target coverage, given that a satisfactory dose level in the target may vary between high and low conflict regions. Thus, Figures \ref{caseB_proton_trade-offs} and \ref{caseB_tomo_trade-offs} show results for Case B in terms of the sample $10^{\text{th}}$ percentile and mean target $\text{D}_{98}$, in addition to criteria \eqref{target_coverage_probability} with $v = 0.98, \gamma = 0.95$. Here, $w_t = 100$ was kept constant, as changing $w_t$ showed little empirical effect on the trade-offs, likely due to the dominance of the maximum dose constraints constraints compared to the remaining dose-limiting terms in the objective. For both modalities, the CTD plan dominates the larger margins for the three presented target coverage criteria. For the margin-based plans, increasing the margin beyond a certain size often has a detrimental effect on the trade-off.

\begin{figure}[!ht]
    \centering
\subfloat[$\text{CTV}_{0.95}$ margin]{%
  \includegraphics[trim={0 0.1cm 0.1cm 0.1cm},clip,width=0.5\textwidth]{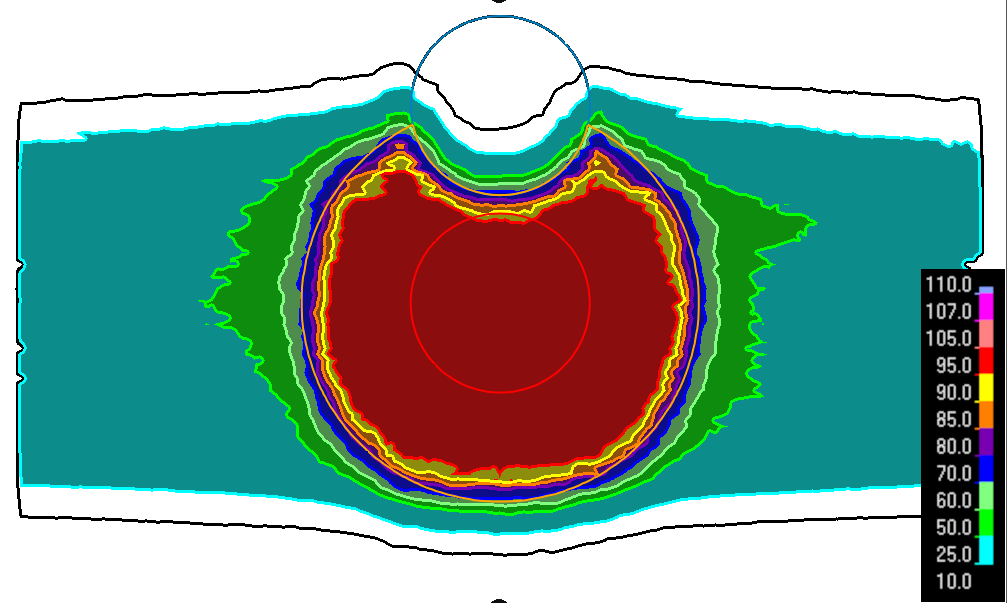}%
  \label{proton_enumeration_095_dose}
}
\subfloat[CTD]{%
  \includegraphics[trim={0 0.1cm 0 0.1cm},clip,width=0.5\textwidth]{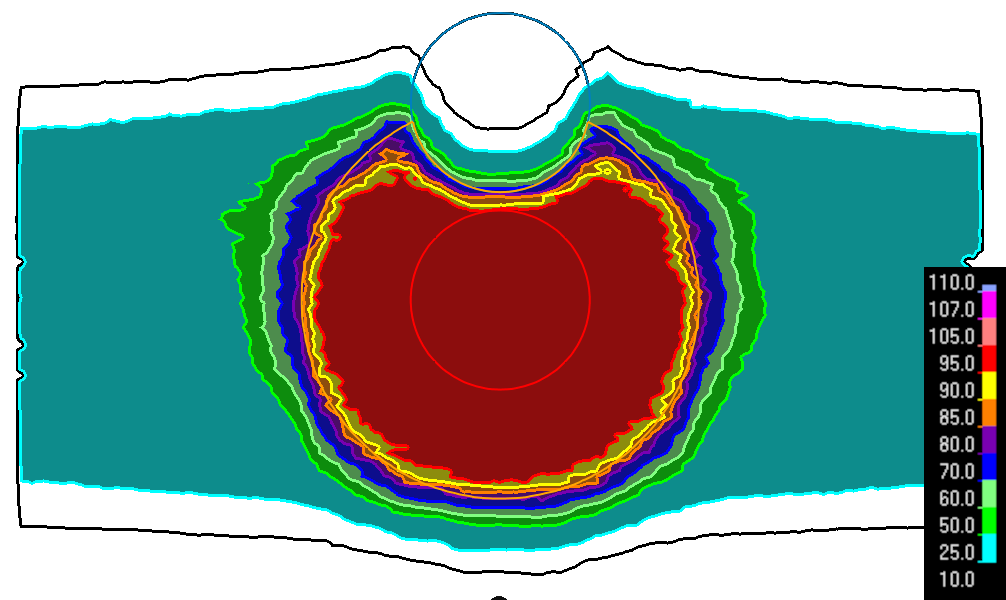}%
  \label{proton_ctd_dose}
}

\subfloat[Dose difference, (b) - (a).]{%
  \includegraphics[clip,width=0.5\textwidth]{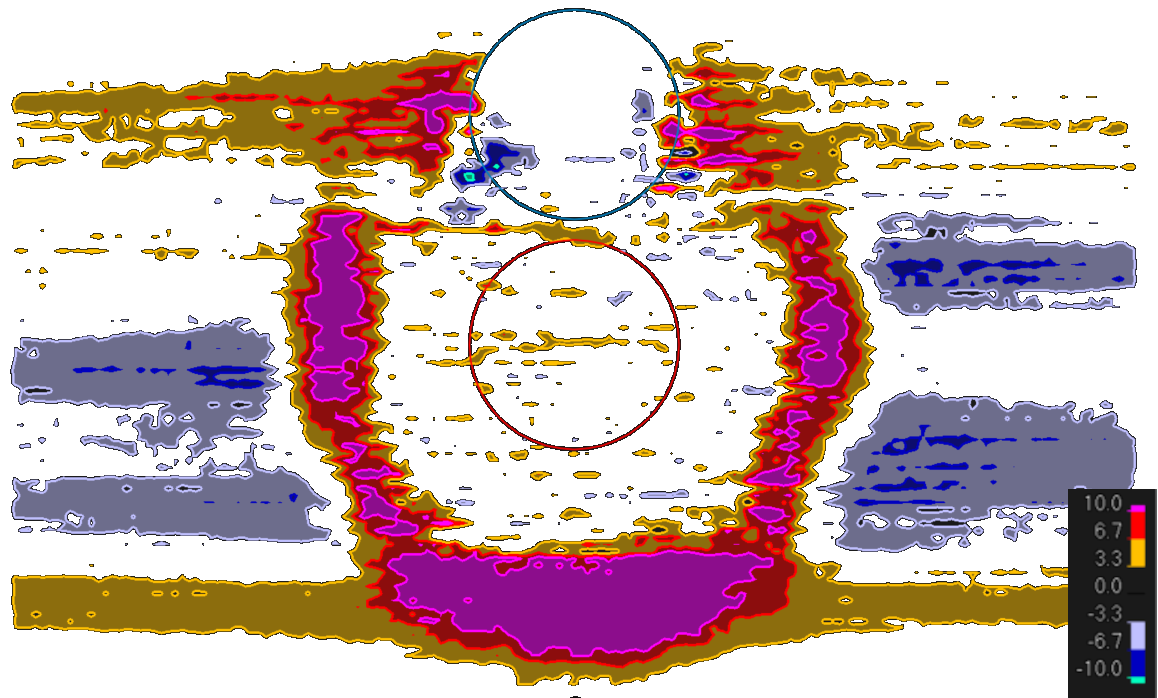}%
  \label{}
}
\subfloat[DVH curves for the External (green) and the OAR (blue). DVH bands between the 0.1- and 0.9-quantiles of the target sample distribution (red), as well as the curve of the median (black).]{%
  \includegraphics[width=0.5\textwidth]{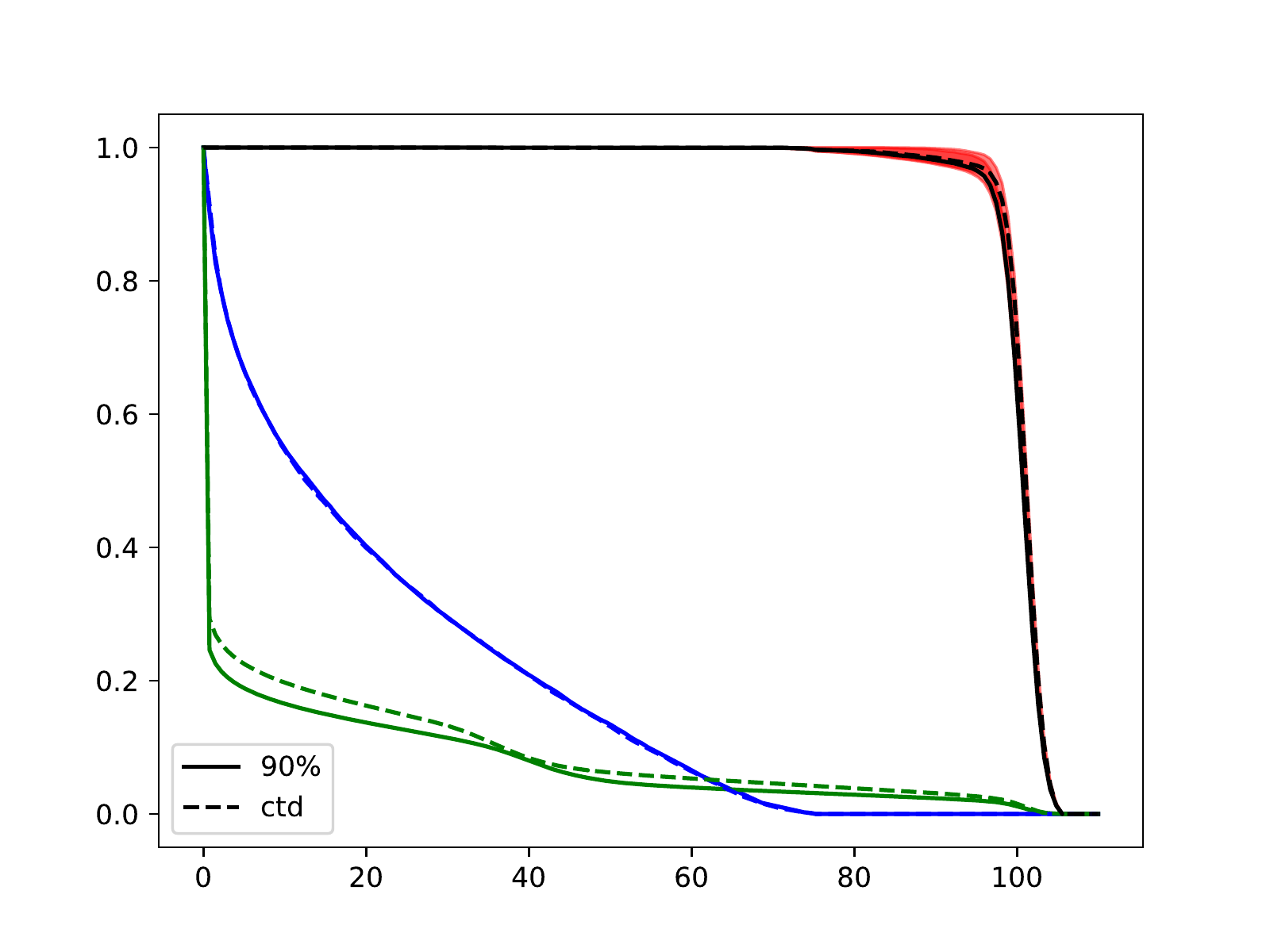}%
  \label{case_b_proton_DVHs}
 }
 \caption{Case B: Dose differences between the considered methods for IMPT.}
 \label{case_b_proton_doses}
\end{figure}

\begin{figure}[!ht]
    \centering
\subfloat[$\text{CTV}_{0.95}$ margin]{%
  \includegraphics[trim={0.1cm 0 0.1cm 0.3cm},clip,width=0.5\textwidth]{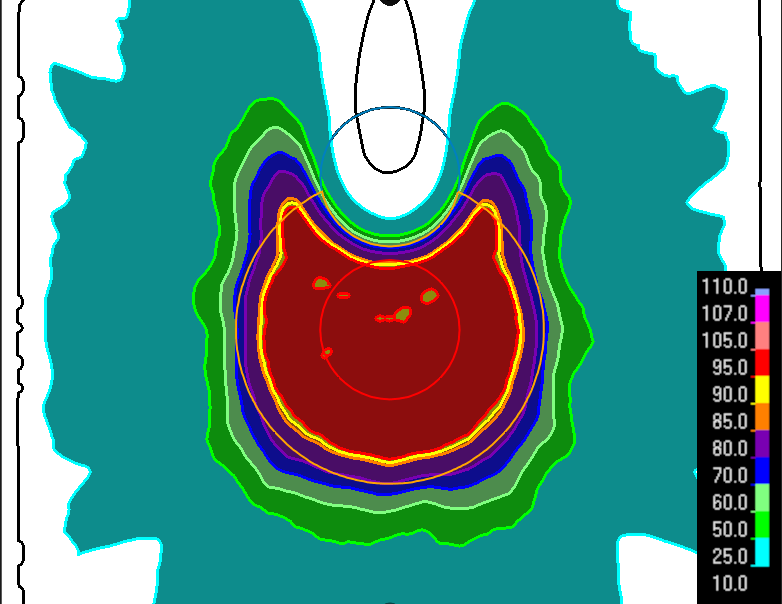}%
  \label{tomo_enumeration_0.95_dose}
}
\subfloat[CTD]{%
  \includegraphics[trim={0.1cm 0 0.1cm 0.15cm},clip,width=0.5\textwidth]{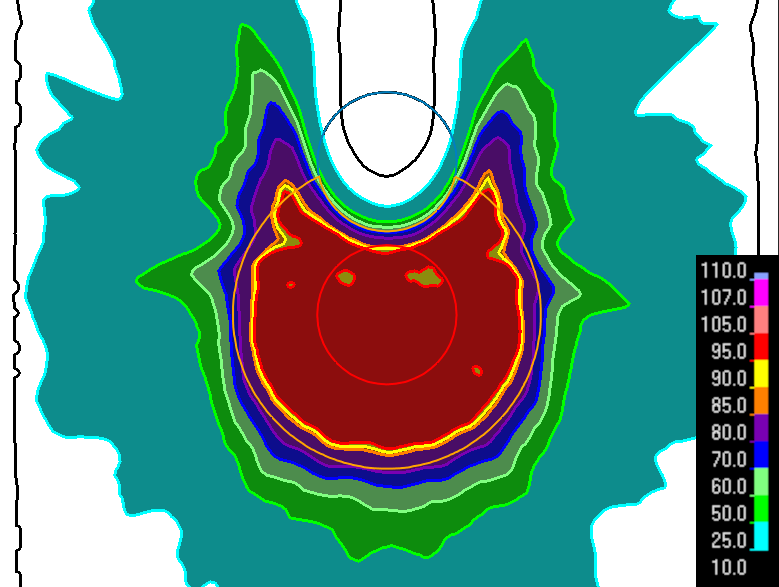}%
  \label{tomo_ctd_dose}
}

\subfloat[Dose difference, (b) - (a).]{%
  \includegraphics[clip,width=0.5\textwidth]{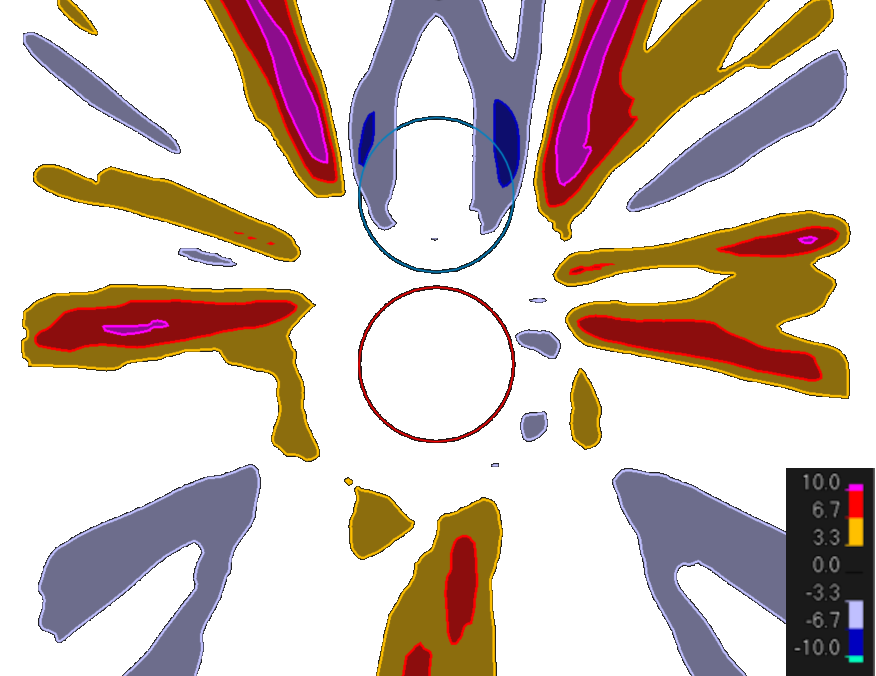}%
  \label{}
}
\subfloat[DVH curves for the External (green) and the OAR (blue). DVH bands between the 0.1- and 0.9-quantiles of the target sample distribution (red), as well as the curve of the median (black).]{%
  \includegraphics[width=0.5\textwidth]{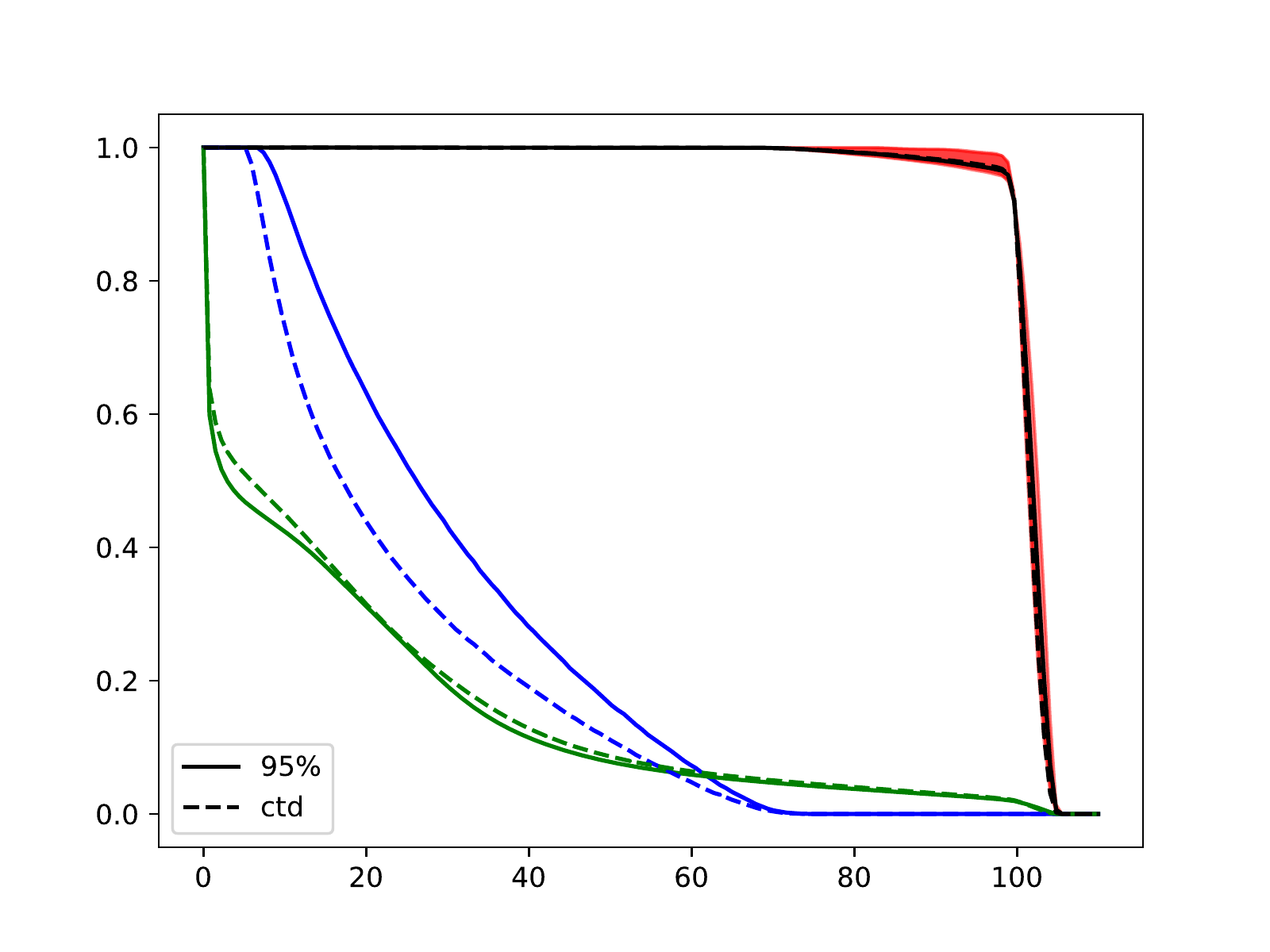}%
  \label{case_b_tomo_DVHs}
 }
 \caption{Case B: Dose differences between the considered methods for tomotherapy.}
 \label{case_b_tomo_doses}
\end{figure}

As for Case A, Figures \ref{case_b_proton_doses} and \ref{case_b_tomo_doses} show doses and DVHs from plans for each of the methods for Case B. For IMPT, the regions in which the doses differ are similar to as in Case A, but the differences are larger in magnitude. The result is a reduced OAR-sparing effect for the CTD-approach, as well as an additional increase in the dose difference near the edge of the CTD. The opposite effect is seen for tomotherapy, for which the OAR-sparing effect of the CTD approach is increased, while the total dose increase is decreased.

\section{Discussion}

In the present work, we have investigated the efficacy of CTD optimization in terms of creating favorable trade-offs between target coverage and over dosage, with respect to the specified criteria. For comparison, conventional margin-based plans were created by enumeration over various CTV-margin sizes. The methods were applied to two phantom geometries to investigate the difference in dose between the resulting plans. In this setting, CTD optimization was shown to be slightly inefficient in the case of the strict near-perfect target coverage criteria from Case A. In Case B, considering a critical, proximal OAR, the plan from the CTD approach dominated those resulting from large CTV margins.

The results for Case A agree with the hypothesis that CTD-based doses are somewhat less conformal, given that low-weight voxels at the edge of the CTD drive the optimal solutions toward giving some but not full dose to the outer regions of the target. Another, related effect of using the CTD approach is that of additional dose to the full patient volume. This result is seen across both Case A and B for both modalities, although in varying magnitude. It should be noted that the increase is not only in the aforementioned regions near the edge of the CTD, but also in low-dose regions, as seen from the DVH curves for the External in Figures \ref{case_a_proton_DVHs}, \ref{case_a_tomo_DVHs}, \ref{case_b_proton_DVHs}, and \ref{case_b_tomo_DVHs}. On the other hand, the CTD approach consistently outperforms the margin-based plans when only the dose to the OAR is considered. Whether this is true in general, or dependent on that the weights to dose-limiting functions to the OAR and the External have been held fixed during the experiments, remains open for investigation.

When introducing CTD-weighted optimization, Shusharina et al. \cite{shusharina_clinical_2018} considered a constraint on mean OAR dose. We think that for the phantom geometries considered in this work, mean and maximum dose constraints have similar effects on the achievable target dose distributions. It is therefore understandable that the results for Case B agree well with those in \cite{shusharina_clinical_2018}.

Here, we present a plausible explaination of the relative advantage of the CTD approach for Case B. For IMPT, using the $\text{CTV}_{0.95}$ margin results in visible under dosage of a small part of the GTV closest to the OAR, as seen in Figure \ref{proton_enumeration_095_dose}. In a sense, the effect is similar to the corresponding case in 1D, illustrated in Figure \ref{1d_results_strict}, in which solutions with respect to large margins achieve a little extra dose to the very edge of the margin by allowing a cold spot closer to the center. We believe that this behavior of the dose is largely a result of the quadratic nature of the dose-promoting minimum dose functions, in which minimization of large deviations from the prescription dose is prioritized. Under these conditions, CTD optimization mitigates this effect by prioritizing central voxels when assigning them greater relative weight in the optimization problem. Although the corresponding result is not seen for tomotherapy in the slice shown in Figure \ref{case_b_tomo_doses}, the probabilities of satisfying $D_{98} > 0.95 \hat{d}$ in Figures \ref{caseB_proton_trade-offs} and \ref{caseB_tomo_trade-offs} suggest that this prioritization of central regions of the target is seen across both modalities.

In the present work, we have disregarded the difficulty in determining the optimal combination of weight and margin. Figure \ref{1d_trade-off} shows that in 1D, varying only $w_t$ in CTD optimization has similar effect as varying the margin size and $w_t$ in the enumeration approach. The trade-offs for Case A suggest that this result also applies for more lenient target coverage criteria in 3D, which is an indication that the CTD approach can produce competitive plans with significantly less computational complexity. The optimal margin is likely to vary with patient geometry, tumor infiltration model, and delivery technique, making it difficult to determine in advance. This difficulty may increase further in even more complex planning situations such as in robust optimization with respect to setup and range uncertainties. In this sense, CTD optimization may be considered as a simple and practical alternative to finding a desirable combination of both margin size and objective weights, since it only requires tuning of the weight on the target.

The results above apply to the considered isotropic tumor infiltration model. For this model, the information contained in the CTD also fully defines the probability distribution over the sample space of potential target volumes. In a more general setting for target delineation uncertainty, the CTD corresponds to each voxel's marginal probability of tumor presence, which specifies the probability of being tumorous without conditioning on the state of any other voxel. In this sense, it is similar to the concept of \textit{spatial probability maps} (SPMs) put forward by van Rooij et al. \cite{van_rooij_using_2021}. In a general setting, marginal probabilites like those from SPMs provide only partial information about the full probability space. Going forward, it may be relevant to investigate the implications of expected value-based optimization (like the CTD approach) with respect to target delineation uncertainty in such a setting.

\appendix

\section{The $\epsilon$-Constraint Method} \label{epsilon-constraint method}

The $\epsilon$-constraint method can be used to find Pareto optimal solutions to the bi-objective problem

\begin{mini}|l|
{x \in \mathcal{X}}{(f_1(x), f_2(x))}{}{},
\label{appendix_biobjective}
\end{mini}

by solving 

\begin{mini}|l|
{x \in \mathcal{X}}{f_1(x)}{}{}
\addConstraint{f_2(x) \leq \epsilon.}
\label{appendix_epsilon}
\end{mini}

It can be shown that a solution $x^*$ is Pareto optimal with respect to problem \eqref{appendix_biobjective} if and only if it is a solution to problem \eqref{appendix_epsilon}, with $\epsilon = f_2(x^*)$. Thus, by fixing $\epsilon$ and solving problem \eqref{appendix_epsilon} for $x^*$, the value of $f_2(x^*)$ can be controlled while ensuring Pareto optimality. For a proof and more information on MCO and the $\epsilon$-constraint method, the reader is referred to Miettinen \cite{miettinen_nonlinear_1998}.

\printbibliography

\end{document}